\documentclass{iopart}

\newcommand{\be} {\begin {equation}}
\newcommand {\ee} {\end {equation}}
\newcommand{\bea} {\begin {eqnarray}\fl}
\newcommand {\eea} {\end {eqnarray}}
\newcommand{\ba}{\begin{array} } 
\newcommand{\ea}{\end{array}}
\newcommand {\ch} {\mathcal{H}}
\newcommand {\cn} {\mathcal{N}}
\newcommand {\cm} {\mathcal{M}}
\newcommand {\cl} {\mathcal{L}}
\newcommand{\nn}{\nonumber\\ }
\newtheorem{theorem}{Theorem}

\newtheorem{corollary}{Corollary}

\begin{document}

\title{An infinite family of superintegrable deformations of the {C}oulomb potential}

\author{Sarah Post$^1$, Pavel Winternitz$^2$}
\address{Centre de recherches math\'ematiques, C.P. 6128 succ. Centre-Ville, Montreal (QC) H3C 3J7, Canada$^1$ }
\address{Centre de recherches math\'ematiques and D\'epartement de math\'ematiques et de statistique, C.P. 6128 succ. Centre-Ville, Montreal (QC) H3C 3J7, Canada$^2$ }
\eads{\mailto{post@CRM.UMontreal.CA}, \mailto{wintern@CRM.UMontreal.CA}}
\begin{abstract}
We introduce a new family of Hamiltonians with a deformed Kepler-Coulomb potential dependent on an indexing parameter $k.$ We show that this family is superintegrable for all rational $k$ and compute the classical trajectories and quantum wave functions. We show that this system is related, via coupling constant metamorphosis, to a family of superintegrable deformations of the harmonic oscillator given by Tremblay, Turbiner and Winternitz. In doing so, we prove that all Hamiltonians with an oscillator term are related by coupling constant metamorphosis to systems with a Kepler-Coulomb term, both on Euclidean space. We also look at the effect of the transformation on the integrals of the motion, the classical trajectories and the wave functions and give the transformed integrals explicitly for the classical system. 
\end{abstract}

\section{Introduction}
The purpose of this article is to introduce an infinite family of classical and quantum systems with the Hamiltonian
\be \label{vdc} V^{DC}_k=-\frac{ Q}{r}+\frac{\alpha k^2}{4r^2\cos^2(\frac{k}{2}\phi)}+\frac{\beta k^2}{4r^2\sin^2(\frac{k}{2}\phi)}\ee 
\be  \label{Hk} \ch_{k}^{DC} =p_1^2+p_2^2+ V^{DC}_k, \qquad  H_{k}^{DC}=-\Delta + V^{DC}_k\ee
where $(r,\phi)$ are polar coordinates with $0\leq \phi \leq \frac{2\pi}{k},\ \vec{p}$ is the linear momentum and $\Delta$ is the Laplacian on 2 dimensional Euclidean space. The superscript denotes that it is a deformed Coulomb potential and the subscript shows the dependence on $k.$ This system was given for $k=1$ in \cite{KMP1996} and $k=2$ in \cite{RTW2009}. 

 We shall show that the classical system is superintegrable for all rational values of k in that it allows two independent integrals of motion, besides the Hamiltonian. Both are polynomial in the momentum, one of second order and the other a higher order polynomial. We show that all bounded classical trajectories of these systems are closed and the motion is periodic. For the quantum system, we show that Schr\"odinger equation is exactly solvable and the energy levels are essentially the same as those of the Coulomb system. We will also show that these systems are related to a family of superintegrable deformations of the harmonic oscillator via coupling constant metamorphosis. 


Superintegrable systems can be classified by the degree of the highest order integral of the motion, excluding the Hamiltonian.  Superintegrable systems of first-order are directly related to Lie groups of point transformations while superintegrable systems of second-order are characterized by separability in multiple coordinate systems. Both types are considered to be well understood \cite{FMSUW, MSVW1967, Dask2006, KKM20061, KKM2007c, DASK2007}. The classification of  higher-order superintegrable systems remains an open problem and has been a subject of much recent activity \cite{ KPRS2002, GW, GRAVEL, EvansVerrier2008, EvansVerrier20082, RTW2008, marquette2009painleve, TW20101, RTW2009, marquette2009super}. 

Most relevant to this paper, is the recent discovery of a family of superintegrable deformations of the harmonic oscillator \cite{TTW2009} indexed by a parameter $k$, referred to by others in subsequent articles as the TTW system.  It can be defined both as a classical or quantum Hamiltonian with potential $V=\omega^2 \rho^2+\alpha k^2\rho^{-2}\sec^2{(k\theta)}+\beta k^2\rho^{-2}\csc^2{(k\theta)}.$ This system has ignited much recent work on its conjectured superintegrability for certain values of $k.$ Specifically, for integer $k,$ it was conjectured to have an independent integral of the motion of order $2k,$ in addition to the second-order integral defining separation of variables. 

 The original authors proved exact-solvability for all $k$ and superintegrability for $k=1,2,3,4.$ Later they demonstrated the periodicity of bounded trajectories in the classical case for rational $k$ \cite{TTW2010}, supporting the conjecture. In recent papers, the classical superintegrability was proven for rational $k$ \cite{KMPTTWClass} and the quantum superintegrability was proven for odd $k$ \cite{QuesneTTWodd}. Most recently, a constructive proof of the superintegrability of the quantum system for rational $k$ was given in \cite{KKM2010quant}.
 
An important tool in the analysis of superintegrable systems is the coupling constant metamorphosis, also referred to as the St\"ackel transform which maps one Hamiltonian to another \cite{HierGrammStackel,BKMstackel}. This transform is particularly useful in the study of integrable and superintegrable systems because it carries an associated mapping of the integrals of the motion. In this paper, we will see that the new Hamiltonian given above is St\"ackel equivalent to the TTW system and so the integrals of the motion, trajectories and wave functions will be intimately linked. 

In Sections 2 and 3, we find the classical trajectories and then the quantum wave functions. In Section 4 we discuss the St\"ackel transform and prove that systems with oscillator terms in the potential are St\"ackel equivalent to systems with Kepler-Coulomb terms and describe the effect of the St\"ackel transform on the trajectories or wave functions  and on the integrals of the motion. In Section 5, we apply these theorems to determine higher-order integrals of the motion for the classical system.

\section{The Classical Trajectories}
We consider the classical trajectories of the system $\ch^{DC}_k$ given by (\ref{vdc}-\ref{Hk}) and prove that the bounded trajectories are closed and the motion is periodic. We follow the procedure in \cite{Goldstein} and separate the action as $S=S_1(r)+S_2(\phi)-Et.$ The Hamilton-Jacobi equation separates as 
\begin{eqnarray}-A&=&r^2(\frac{\partial S_1}{\partial r})^2-Qr-Er^2\\
-A&=&-\left((\frac{\partial S_2}{\partial \phi })^2+\frac{\alpha k^2}{4\cos^2(\frac{k}{2}\phi)}+\frac{\beta k^2}{4\sin^2(\frac{k}{2}\phi)}\right)\eea
and for $S_1$ and $S_2,$ we have 
\[\fl S_1(r)=\int\frac{1}{r}\sqrt{Er^2+Qr-A}dr, \quad S_2(\phi)=\int\sqrt{A- \frac{\alpha k^2}{4\cos^2(\frac{k}{2}\phi)}-\frac{\beta k^2}{4\sin^2(\frac{k}{2}\phi)}}.\]
The trajectories will then satisfy
\be\label{Hkdeteq} \frac{\partial S}{\partial E}=\frac{\partial S_1}{\partial E}-t=\delta_1, \quad \frac{\partial S}{\partial A}=\frac{\partial S_1}{\partial A}+\frac{\partial S_2}{\partial A}=\delta_2.\ee
It remains to integrate these equations under the condition that the motion be bounded, 
 \be\label{rbounds} 0\leq r_1\leq r\leq r_2, \quad Er_i^2+Qr_i-A=0,\quad i=1,2.\ee
Also, we have the requirement
\be \fl\label{phibounds} 0\leq u_1\leq cos^2(\frac{k}{2}\phi)\leq u_2\leq1, \quad  -Au_i^2+(A -\beta (\frac{k}2)^2+\alpha (\frac{k}2)^2)u_i-\alpha(\frac{k}2)^2=0.\ee
These conditions give restrictions on the choices of parameters. We summarize these in Table \ref{table} below.

\begin{table}[h!b!p!]
\caption{Parameter restrictions for bounded trajectories}
\label{table}
\begin{tabular}{|c|c||c|c|} \hline
Restriction& Effect &Restriction & Effect\\ 
\hline
$D_1\equiv Q^2+4AE>0$& $r_i$ real &$D_2\equiv (A-\frac{k^2}{4}(\beta+\alpha))^2-\frac{\alpha\beta k^4}{4}>0$& $u_i$ real\\ 
$Q>0$& $0<r_2$&$A-\frac{k^2}{4}|\beta-\alpha|>0$& $u_2>0$\\
$A>0$& $0<r_1$ and &$\beta>0$ & $u_1>0$\\
& $u_1<\cos^2(\frac{k}{2}\phi)<u_2$&&\\
$E<0$& $r_1<r<r_2$&$\alpha>0$ &$u_2<1$\\ 
\hline
\end{tabular} \end{table}

Under the restrictions given in the table above, the solutions for (\ref{Hkdeteq}) are 
\be \fl \label{deteq1} 0= \frac{-2Er-Q}{\sqrt{D_1}}+\sin\left(\frac{4(-E)^{3/2}}{Q}(t+\delta_1)+\frac{2\sqrt{-E}}{a}\sqrt{Er^2+Q r-A}\right)\ee
\be\fl \label{deteq2} 0=  -\delta_2+\frac{1}{2\sqrt{A}}\arcsin(\frac{2A-Qr}{r\sqrt{D_1}})+\frac{1}{2k\sqrt{A}}\arcsin(\frac{2A\sin^2(\frac{k}{2}\phi)-A+\frac{k^2}{4}(\beta-\alpha)}{\sqrt{D_2}})\ee
and hence $r$ has the period of $\frac{ Q \pi}{2(-E)^{3/2}}$ in $t.$ These trajectories are also periodic in $\phi$ for rational $k=c/d$ with $c,d$ integer. Rewriting (\ref{deteq2}), 
 \be \fl 0 =-2\sqrt{A}c \delta_2+\frac{c+d}{2}\pi-c\arccos(\frac{2A-Qr}{r\sqrt{D_1}})-d\arccos(\frac{2A\sin^2(\frac{k}{2}\phi)-A+\frac{k^2}{4}(\beta-\alpha)}{\sqrt{D_2}}) \ee
and using the Chebyshev polynomials defined as, 
\be \label{Cheby} T_n(x)=\cos\left(n\arccos(x)\right), \qquad U_n(x)=\frac{\sin\left((n+1)\arccos(x)\right)}{\sin\arccos x}\ee
we obtain, 
\bea \label{rphi} 0=- T_c(\frac{2A-Qr}{r\sqrt{D_1}})+\cos(C)T_d\left(\frac{2A\sin^2(\frac{k}{2}\phi)-A+\frac{k^2}{4}(\beta-\alpha)}{\sqrt{D_2}}\right)\\
\fl \quad +\sin(C)U_{d-1}\left(\frac{2A\sin^2(\frac{k}{2}\phi)-A+\frac{k^2}{4}(\beta-\alpha)}{\sqrt{D_2}}\right)\sqrt{1-\left(\frac{2A\sin^2(\frac{k}{2}\phi)-A+\frac{k^2}{4}(\beta-\alpha)}{\sqrt{D_2}}\right)^2}\nonumber\eea
where $C=(-2\sqrt{A}p\delta_2+(c+d)/2\pi).$ Under the given restrictions on the parameters, the implicit function for $r=r(\phi)$ given by (\ref{rphi}) is well defined and periodic, with period $\tau=\frac{2}{k}\pi$ in $\phi.$

We note that these trajectories and their determining equations bear a striking resemblance to those obtained for the TTW system \cite{TTW2010}. In fact, the implicit function determining $r=r(\phi)$ is identical under a change of variables and parameters. We shall see in a later section why this is so. 
  
\section{Eigenfunctions for the Quantum System}
In this section, we solve for the wave functions of the quantum system $H_k^{DC}$ given by (\ref{vdc}-\ref{Hk}) and show that the system is exactly solvable. That is, its energy values can be calculated algebraically and the eigenfunctions can be realized as polynomials modulo a gauge function \cite{Turbiner1988,TempTW}. 

We assume a solution $\Psi=R(r)S(\phi)$ and separate the equation $H^{DC}_{k}\Psi=E\Psi$ as
\be\label{Req} \left(-\partial_r^2-\frac1r\partial_r+\frac{A}{r^2}-\frac{Q}{r}-E\right)R(r)=0.\ee
\be\label{Aeq} \left(-\partial_\phi^2+\frac{\alpha k^2}{4\cos^2(\frac{k\phi}2)}+\frac{\beta k^2}{4\sin^2(\frac{k\phi}2)}-A\right)S(\phi)=0.\ee

We look for solutions which can be written as gauge transformations of a polynomial, a characteristic of exact solvability. In order to obtain such a form of solutions, we require that $\alpha$ and $\beta$ be greater that $-1/4$ and rewrite $\alpha=a(a-1), \beta=b(b-1).$ 

If we take a gauge transformation with gauge  $G_1=r^{\sqrt{A}}e^{2r\sqrt{-E}}$, the transformed radial equation (\ref{Req}) will have polynomial solutions if we restrict to quantized values of the energy $E=-Q^2(2n+1+2\sqrt{A})^{-2}$. If we make a gauge transformation with $G_2=\cos(\frac{k}{2}\phi)^{a}\sin(\frac{k}{2}\phi)^{b}$ then the transformed angular equation (\ref{Aeq}) will have polynomial solutions if we restrict to quantized values of the parameter $A=k^2(2m+a+b)^2/4.$ 

A set of solutions for the Schr\"odinger equation is given by Jacobi polynomial multiplied by Laguerre polynomials
\be\Psi=G_1 G_2 L_{n}^{2\sqrt{A}}\left(2r\sqrt{-E}\right)P_m^{a-\frac12,b-\frac12}\left(-\cos(k\phi)\right) \ee
with energy
\be E=\frac{-Q^2}{\left(2(n+km)+1+ka+kb\right)^2} .\ee
For a given rational $k=c/d$ the energy levels are indexed by an integer $N=dn+cm$ and their degeneracy is $D=[dN/c]+1.$ This coincides with the degeneracy of an anisotropic oscillator with frequency ratio $c/d.$


These wave functions are in agreement with the solutions for the $k=1$ case previously analyzed \cite{KKM1996}. The quantum system is indeed exactly solvable and we recover the requirements $A>0, E<0 $ with a slight relaxing in the restrictions on the parameters $\alpha, \beta$ which are both required to be greater that $-1/4$ instead of positive as in the classical case. Here we see again the relation between these eigenfunctions and those of the TTW system which differ by the same change of variables and parameters as the classical trajectories. In the next section we shall see why this is the case as we prove some theorems which are directly relevant to our systems. 

\section{ Coupling Constant Metamorphosis of Hamiltonians separable in polar coordinates with a harmonic oscillator term in the potential}
Consider a classical Hamiltonian $\ch =\hat{\ch}-\tilde{E} U,$ where $\hat{\ch}$ includes the kinetic energy and part of the potential independent of the coupling constant ($-\tilde{E}$).  We can then write the Hamilton-Jacobi equation $\ch=E$ and solve for $\tilde{E}$ to obtain a new Hamilton-Jacobi equation $\tilde{\ch}\equiv U^{-1}(\hat{\ch}-E)=\tilde{E}.$ For $\ch$ and $\tilde{\ch}$ the role of coupling constant and the energy are interchanged. The quantum version follows by direct analogy. Such a transform is called coupling constant metamorphosis or St\"ackel transform. A remarkable characteristic of coupling constant metamorphosis is that there is an associated mapping of the integrals of the motion to the new system. Such mapping was given first for classical integrable Hamiltonians and later extended to quantum systems with second-order integrals \cite{HierGrammStackel,BKMstackel}. More recently, the quantum transform was extended to higher order constants of the motion \cite{KMPostStackel}. 





The theorems which define the transforms are given below with their proofs. 
\begin{theorem}\label{ccmclass} Given a classical Hamiltonian $\ch=\hat{\ch}-\tilde{E}U,$ where $\hat{\ch}$ is independent of the arbitrary parameter $\tilde{E},$ with an integral of the motion $\cl(\tilde{E}).$ If we define the St\"ackel transform of $\ch$ and $\cl$ as  $\tilde{\ch}\equiv U^{-1}(\hat{\ch}-E)$ and $\tilde{\cl}\equiv \cl(\tilde{H})$ then $\tilde{\cl}$ is an integral of the motion for $\tilde{\ch}.$ 
\end{theorem}
To prove this, we use an identity for Poisson brackets given by
$$ \lbrace F(p_i, q_j, f(p_i,q_j)),G\rbrace=\lbrace F(p_i, q_j, \tau),G\rbrace|_{\tau=f(p_i,q_j)}+\frac{\partial F(p_i,q_j, \tau)}{\partial \tau}|_{\tau=f(p_i,q_j)}\lbrace f(p_i,q_j),G\rbrace$$
where $p_i, q_j$ are the conjugate position and momenta and $\tau$ is a parameter. With this, we compute, 
\begin{eqnarray} \lbrace \tilde{\ch},\tilde{\cl}\rbrace&=& \lbrace \frac{1}{U}(\ch+\tilde{E}U-E),\cl |_{\tilde{E}
 =\tilde{\ch}}\rbrace\nn
 &=&\lbrace \frac{1}{U}(\ch+\tilde{E}U-E),\cl \rbrace|_{\tilde{E}=\tilde{\ch}}+\partial_{\tilde{E}}\cl(\tilde{E})|_{\tilde{E}=\tilde{\ch}}\lbrace \tilde{\ch},\tilde{\ch}\rbrace \nn
&=&-\lbrace U,\cl\rbrace\frac{1}{U^2}(\ch-E)|_{\tilde{E}=\tilde{\ch}}\nonumber\eea
and since $\ch|_{\tilde{E}=\tilde{\ch}}=E,$ we see that  $\lbrace \tilde{\ch},\tilde{\cl}\rbrace=0$ and we have proved the theorem. 

There is an associated theorem for quantum systems though we must make a further assumption about the form of the integral of motion in order to get a well defined integral. 

\begin{theorem} Given a quantum Hamiltonian $H=\hat{H}-\tilde{E}U,$ where $\hat{H}$ is independent of the arbitrary parameter $\tilde{E},$ with an integral of the motion $L=\sum_{j=0}^{[\frac{n}{2}]}K_{N-2j}\tilde{E}^j, $ where $K_{i}$ have degree $i$ as differential operators.  If we define the St\"ackel transform of $H$ and $L$ as $\tilde{H}=U^{-1}(\hat{H}-E)$ and $\tilde{L}=\sum_{j=0}^{[\frac{n}{2}]}K_{N-2j}\tilde{H}^j, $ then $[\tilde{H}, \tilde{L}]=0.$ 

Furthermore, if  $H$ is self-adjoint and $L$ is self or skew adjoint, depending on the parity of $N,$ with respect to $d\mu$ then $\tilde{H}$ will be self-adjoint and $\tilde{L}$ will have the same parity as $L$ with respect to the metric $Ud\mu.$ 
\end{theorem}

The proof of this theorem uses the fact that the $K_i's$ do not depend on $\tilde{E}$ to show that for all integer $j,$ we have 
\be\fl \label{recursion} [L,H]=[\sum_{j=0}^{[\frac{N}{2}]}K_{N-2j}\tilde{E}^{j},\hat{H}+\tilde{E}U]=0 \iff  [ K_{N-2j},\hat{H}]+[K_{N-2j+2},U]=0, \ee 
and 
\be \label{adjoint} \fl \int Lfgd\mu=(-1)^N\int fLgd\mu\iff \int K_{N-2j}fgd\mu=(-1)^N\int f K_{N-2j}gd\mu,\ee
where we have extended the $K_i$ to all integers by setting $K_i=0$ if $i<0$ or $i>N.$
We then use $ [K_{N-2j},\tilde{H}]=U^{-1}[K_{N-2j},\hat{H}]-U^{-1}[K_{N-2j},U]\tilde{H}$ to compute
\be[\tilde{L},\tilde{H}]=\sum_j\left( [ K_{N-2j},\hat{H}]+[K_{N-2j+2},U]\right)\tilde{H}^j=0\ee
where the last equality follows by (\ref{recursion}). Similarly, (\ref{adjoint}) implies
\be \int \sum_j K_{n-2j}\tilde{H}^jfgUd\mu=\sum_j (-1)^N\int f(\tilde{H}^jU^{-1}K_{N-2j}Ug)Ud\mu\ee
while (\ref{recursion}) gives $ \sum_j \tilde{H}^jU^{-1}K_{N-2j}U=\sum_j K_{N-2j}\tilde{H}^j.$
Hence, the equality $\int \tilde{L}fgUd\mu=(-1)^N\int f\tilde{L}gUd\mu$ holds and we have proved the theorem. 

It is important to note that the St\"ackel transform usually maps one system to another on a different ambient manifold since there is a conformal transform of the metric. However, for certain forms of $U,$ the system is still on Euclidean space, though with a different choice of variables; this will be the case for $U=\rho^2.$ If we assume further that the Hamiltonian is separable in polar coordinates, we have the following theorems. 

\begin{theorem}\label{ccmclass1} Given a classical Hamiltonian $\ch$ in $4$-dimensional phase space,  separable in polar coordinates and with a term in the potential corresponding to an isotropic oscillator, i.e. of the form 
\be \label{hclass} \ch(\rho,\theta)=p_\rho^2 +\frac1 {\rho^2}p_\theta^2-\tilde{E}\rho^2+f_1(\rho)+\frac{1}{\rho^2}f_2(\theta)\ee where $f_1(r)$ and $f_2(\theta)$ are independent of $\tilde{E}$. The St\"ackel transform of $\ch$ is again on Euclidean space and given by 
\be \label{htclass} \tilde{\ch}(r,\phi)=p_r^2+\frac{1}{r^2}p_\phi^2-\frac{E}{2r}+\frac{1}{2r}f_1(\sqrt{2r})+\frac{1}{4r^2}f_2(\frac{\phi}2)\ee \end{theorem}
This theorem can be directly verified by taking the St\"ackel transform of $\ch$ as given in Theorem 1
\[ \tilde{\ch}(\rho,\theta)=\frac{1}{\rho^2}\left(p_\rho^2 +\frac1 {\rho^2}p_\theta^2+f_1(\rho)+\frac{1}{\rho^2}f_2(\theta)-E\right)\]
and making the change of variables  $r=\rho^2 /2, \phi=2\theta.$ 
We have the following immediate result.
\begin{corollary}
The Hamiltonian given by (\ref{htclass}) is separable in polar coordinates with an associated integral of the motion $\cl_1=p_\phi^2+\frac{1}{4}f_2(\frac{\phi}2).$ Furthermore, if $\ch$ has an additional integral of the motion then so will $\tilde{\ch}$.
\end{corollary}
The first assertion can be observed from the Hamiltonian and both assertions are results of Theorem 1. 
There is also a constructive relation between the trajectories of the two systems. 
\begin{theorem}\label{ccmclasstrajectories}
 If $\ch(\rho,\theta)$ as given in (\ref{hclass}) has trajectories $(\rho(t),\theta(t))$ which satisfy $\delta_1=F_1(\rho)-t, \delta_2=F_2(\rho,\theta) $ then trajectories for $\tilde{\ch}(r,\phi)$ will satisfy

\be \delta_1= \frac{d}{d\tilde{E}}\int F_1(\sqrt{2r})dE-t,\quad  \delta_2=F_2(\sqrt{2r},\frac{\phi}{2}),\ee
\end{theorem}

We prove this by following the same procedure as in the previous section to solve for the trajectories.  We separate the  action as $S=S_1(\rho)+S_2(\theta)-Et$ for $\ch$ and $\tilde{S}=\tilde{S_1}(\sqrt{2r})+\tilde{S_2}(\frac{\phi}{2})-\tilde{E}t$ for $\tilde{\ch}.$ By construction, the Hamilton-Jacobi equations are identical, the only differences being that the energy for the system $\ch$ is $E$ while $\tilde{E}$ is the energy of the system $\tilde{\ch}$. Thus, we have $S_1(\rho)=\tilde{S_1}(\rho)$ and $S_2(\theta)=\tilde{S_2}(\theta).$ Therefore, the trajectories for $\tilde{\ch}$ must satisfy
\be \left(\frac{\partial S_1}{\partial \tilde{E}}\right)-t=\delta_1,\qquad \frac{\partial S_1}{\partial A}+\frac{\partial S_2}{\partial A}=\delta_2.\ee
Hence, if the trajectories for $\ch,\  (\rho(t), \theta(t)),$ satisfy $\delta_1=F_1(\rho)-t, \ \delta_2=F_2(\rho,\theta)$ then 
\[ F_1(\rho)= \frac{\partial S_1}{\partial E},\quad F_2(\rho,\theta)=\frac{\partial S_1}{\partial A}+\frac{\partial S_2}{\partial A}\]
 and so the trajectories $\tilde{\ch},(r(t),\phi(t)),$ must satisfy
\[ \delta_1= \frac{\partial}{\partial \tilde{E}}\int F_1(\sqrt{2r})dE-t, \quad \delta_2=F_2(\sqrt{2r},\frac{\phi}{2}).\]
We have a similar result for the quantum system. 
\begin{theorem}\label{ccmquant} Given a quantum Hamiltonian $H(\rho,\theta)$ in 2 dimensions, separable in polar coordinates and  with a term in the potential corresponding to an isotropic oscillator, i.e. of the form 
\be \label{hquant} H(\rho,\theta)=-\frac{1}{\rho}\partial_\rho(\rho\partial_\rho) -\frac1 {\rho^2}\partial_\theta^2-\tilde{E}\rho^2+f_1(\rho)+\frac{1}{\rho^2}f_2(\theta)\ee
where $f_1(\rho)$ and $f_2(\theta)$ are independent of $\tilde{E}$. The St\"ackel transform of $H(\rho,\theta)$ is again on Euclidean space and given by 
\be  \label{htquant} \tilde{H}(r, \phi)=-\frac{1}{r}\partial_r(r\partial_r)-\frac{1}{r^2}\partial_\phi^2-\frac{E}{2r}+\frac{1}{2r}f_1(\sqrt{2r})+\frac{1}{4r^2}f_2(\frac{\phi}{2})\ee
\end{theorem}
This theorem can be directly verified by taking the St\"ackel transform of $H$ as given in Theorem 2
\[ \tilde{H}(\rho,\theta)=\frac{1}{\rho^2}\left(-\frac{1}{\rho}\partial_\rho(\rho\partial_\rho) -\frac1 {\rho^2}\partial_\theta^2-E+f_1(\rho)+\frac{1}{\rho^2}f_2(\theta)\right)\]
and making a change of variables of  $r=\rho^2 /2, \phi=2\theta.$ 
We have the following immediate result.
\begin{corollary}
The Hamiltonian given by (\ref{htquant}) is separable in polar coordinates with an associated integral of the motion $L_1=-\partial_\phi^2+\frac{1}{4}f_2(\frac{\phi}2).$ Furthermore, if $H$ has an additional integral of the motion of the form given in Theorem 2 then so will $\tilde{H}$.
\end{corollary}
In addition, the solutions to the Schr\"odinger equation for $H$ are also solutions to the Schr\"odinger equation for $\tilde{H}$ since, by construction, $(\tilde{H}(\rho,\theta)-\tilde{E})=\frac{1}{\rho^2}(H(\rho,\theta)-E).$
\begin{theorem} If $\Psi(\rho,\theta)$ is a solution to $H(\rho,\theta)\Psi(\rho,\theta)=E\Psi(\rho,\theta)$ then $\Psi(\sqrt{2r},\frac{\phi}{2})$ will be a solution to $\tilde{H}(r,\phi)\Psi(\sqrt{2r},\frac{\phi}{2})=\tilde{E}\Psi(\sqrt{2r},\frac{\phi}{2}).$
\end{theorem}

If we consider the case where $\tilde{E}=-\omega^2,$ $f_1(r)=0$ and $f_2(\theta)=\alpha k^2\rho^{-2}\sec^2(k\theta)+\beta k^2\rho^{-2}\csc^2(k\theta)$ then the original Hamiltonians $\ch, H$ corresponds to the TTW system, $\ch^{TTW}_k,H^{TTW}_k$ \cite{TTW2009}. Furthermore, if we take $E=Q/2$ then the transformed Hamiltonians $\tilde{\ch}^{TTW}_k, \ \tilde{H}^{TTW}_k$ will coincide with $\ch_k^{DC}, \ H_k^{DC}$ so we can use the superintegrability of the TTW system to prove the superintegrability of the deformed Coulomb system.

\section{The Higher-Order Integral of the Motion}
We can directly apply the above theorems to the TTW system to obtain an additional integral of the motion for $\ch^{DC}_k$ by taking the St\"ackel transform of the higher-order constant of the motion of the TTW system. The existence of constants of the motion for the classical TTW Hamiltonian were recently proven for all rational $k$ \cite{KMPTTWClass}. We would like to see the mapping of these integrals of the motion under the St\"ackel transform. In particular, we want to verify that they will remain polynomial in the momenta. To do this, we derive an explicit expression for the integral and show that these are polynomials in the parameter $\omega^2.$   

We begin by finding an explicit expression for the constants of motion. First, we redefine $\rho=e^{R}$ in order to express the Hamiltonian and the integral of the motion as  
\be \ch^{TTW}_k =e^{-2R}\left(p_R^2+\omega^2 e^{4R}+\cl_1\right)\ee
\be  \cl_1=p_{\theta}^2+\frac{\alpha k^2}{\cos^2(k\theta)}+\frac{\beta k^2}{\sin^2(k\theta)}\ee
The auxiliary functions $\cm,\cn$ given in \cite{KMPTTWClass} can be written in a modified form as  
$$  \cm=\frac{1}{4\sqrt{\cl_1}} arccos\left(\frac{B_x}{\sqrt{B_y^2+B_x^2}}\right),\quad  \cn=\frac{1}{4k\sqrt{\cl_1}}arccos\left(\frac{A_x}{\sqrt{A_x^2+A_y^2}}\right)$$
\begin{eqnarray}A_y=\cl_1\cos(2k\theta)-\alpha k^2+\beta k^2, &\qquad& A_x=\sqrt{\cl_1}\sin(2k\theta)p_\theta,\nn
 B_y=2\cl_1e^{-2R}-\ch^{TTW}_k, &\qquad & B_x=2\sqrt{\cl_1}e^{-2R}p_R.\nonumber \end{eqnarray}
 Notice that, because the quantities 
 \begin{eqnarray} A_x^2+A_y^2=(\cl_1-(\alpha+\beta)k^2)^2-4k^4\alpha\beta,\nn
 B_x^2+B_y^2=(\ch^{TTW}_k)^2-4\omega^2\cl_1 .\end{eqnarray}
 depend only on $\ch^{TTW} _k,\ \cl _1 $ and parameters, we can always multiply an integral of the motion by a function of these and it will still Poisson commute with $\ch^{TTW} _k$.
 
 Since the functions $\cm$ and $\cn$ satisfy $\lbrace\cm,\ch^{TTW}_k\rbrace=\lbrace \cn,\ch^{TTW}_k \rbrace=e^{-2R},\cm-\cn$ will be an integral of the motion, though not polynomial in the momenta. However, for rational $k=c/d$ with $c,d$ integer, the integral can be put into a form so that it is polynomial in the momenta. One such integral is
\be \label{L2} \cl_2^{(\sin)}\equiv \left(\sqrt{B_x^2+B_y^2}\right)^{c}\left(\sqrt{A_x^2+A_y^2}\right)^{d}\frac{\sin \left(4c\sqrt{\cl_1}(\cm-\cn)\right)}{\sqrt{\cl_1}^{\delta_{c+d-1}}}\ee
where $\delta_i$ is $0$ if $i$ is even and $1$ when $i$ is odd.  To show that this integral is polynomial in the momenta, we use  the identities for Chebyshev polynomials (\ref{Cheby}) to rewrite $\cl_2$ as
\bea\cl_2^{(\sin)}=\left(\sqrt{B_x^2+B_y^2}\right)^{c}\left(\sqrt{A_x^2+A_y^2}\right)^{d}\Bigg[
\frac{B_y} {\sqrt{B_x^2+B_y^2}}U_{c-1}\left(\frac{B_x} {\sqrt{B_x^2+B_y^2}}\right)T_d\left(\frac{A_x}{\sqrt{A_x^2+A_y^2}}\right)\nn
-\frac{A_y} {\sqrt{A_x^2+A_y^2}}T_c\left(\frac{B_x} {\sqrt{B_x^2+B_y^2}}\right)U_{d-1}\left(\frac{A_x} {\sqrt{A_x^2+A_y^2}}\right) \Bigg]\frac{1}{\sqrt{\cl_1}^{\delta_{c+d-1}}}\eea
which can be expanded as
\bea\cl_2^{(\sin)}=\frac{1}{\sqrt{\cl_1}^{\delta_{c+d-1}}}\Bigg[\left(\sum_{m=0}^{\left[\frac{c-1}{2}\right]}(\ba{c} c\\ 2m+1\ea)(-1)^mB_x^{c-2m-1}B_y^{2m+1}\right) \left(\sum_{m=0}^{\left[\frac{d}{2}\right]}(\ba{c} d\\ 2m\ea)(-1)^mA_x^{d-2m}A_y^{2m}\right)\nn
\fl \qquad\label{L3}-\left(\sum_{m=0}^{\left[\frac{d-1}{2}\right]}(\ba{c} d\\ 2m+1\ea)(-1)^m A_x^{d-2m-1} A_y^{2m+1}\right) \left(\sum_{m=0}^{\left[\frac{c}{2}\right]}(\ba{c} c\\ 2m\ea)(-1)^mB_x^{c-2m}B_y^{2m}\right)\Bigg].\eea
This is polynomial in the momenta because of the definition of the $A's $ and $B's$ and because of the parity of the Chebyshev polynomials. 

As demonstrated in the orginal paper, there is also an integral of the motion obtained by taking cosine instead. It can be written as
\be \cl_2^{(\cos)}\equiv \left(\sqrt{B_x^2+B_y^2}\right)^{c}\left(\sqrt{A_x^2+A_y^2}\right)^{d}\frac{\cos \left(4c\sqrt{\cl_1}(\cm-\cn)\right)}{\sqrt{\cl_1}^{\delta_{c+d}}}.\ee
Because of the power of $\sqrt{\cl_1}$ in the denominator, the degrees of $\cl_2^{(\sin)}$ and $\cl_2^{(\cos)}$ will differ by 1, with the lowest degree being $2(c+d)-1.$  In the case of $k$ integer, we do not get a symmetry operator of degree $2k$ but instead one of degree at least $2k+1.$ In this case, the operator of lowest degree, namely $2k+1$ will be $\cl_2^{(\cos)}$ for even $k$ and $\cl_2^{(\sin)}$ for odd $k$. 
Thus, while it has been explicitly proven that the systems are classically superintegrable for rational values of $k,$ there is still no proof that the integrals of the motion can be written as polynomials in the momenta of degree $2k$ for integer $k.$ We conjecture that there is such an integral $\cl_3$ and it can be related to the integrals given above by $\cl_2^{(\mu)}=\lbrace \cl_3,\cl_1\rbrace, $ for $\mu=\sin, \cos$ depending on the parity of $k$. We have verified the conjecture for $k=1, 2.$

From the equations above, we see explicitly that the integrals of motion are polynomial in the parameter $\omega^2$ and hence coupling constant metamorphosis will map them to integrals of the motion for $\tilde{\ch}^{TTW}_k$ which are still polynomial in the momenta. To determine such integrals we must replace $\omega^2=-\tilde{\ch}^{TTW}_k$ and perform the requisite change of variables. Hence, the constants of the motion for Hamiltonian $ \tilde{\ch}^{TTW}_k$ will be \be\tilde{\cl_2}^{(\mu)}(r,\phi)=\cl_2^{(\mu)}(\sqrt{2r},\phi/2)|_{\omega^2=-\tilde{\ch}^{TTW}_k}, \qquad \mu=\sin,\cos.\ee
 Finally, $\ch^{DC}_k$ is related to $\tilde{\ch}^{TTW}_k ,$ by the parameter change $E=Q/2$ and so the integrals will need to under go the same transform. The existence of these additional integrals of motion proves that the classical system $\ch^{DC}_k$ is superintegrable.

\section{Conclusion}
In this paper, we have presented a new infinite family of superintegrable systems associated with a deformation of the Coulomb potential. We have shown that the bounded trajectories for the classical system are periodic and that the quantum system is exactly solveable.  We have also shown that deformed Coulomb system is St\"ackel equivalent to the TTW system and used the classical St\"ackel transform to show that the new Hamiltonian is classically superintegrable. We mention that Bertrand's theorem \cite{Goldstein, Bertrand} (valid in $n$ dimensions) tells us that the only spherically symmetric potentials for which all classical bounded trajectories are closed are the harmonic oscillator and the Coulomb-Kepler potential. The potentials $\alpha r^{-1}$ and $\alpha r^2$ are also the only two spherically symmetric potentials which are maximally superintegrable. We now see that at least for $n=2$ both of these systems can be deformed into infinite families of superintegrable systems by adding a symmetry breaking term as in (\ref{vdc}) and that the two families are related via coupling constant metamorphosis. 

As a tool in our analysis, we have proven some general theorems about coupling constant metamorphosis and its action on a class of Hamiltonians, of which the TTW is an example. Though the Hamiltonians $\ch^{DC}_k, H^{DC}_k$ are certainly novel systems, there is a direct relation induced by coupling constant metamorphosis between not only the integrals of the motion but also the trajectories and wave functions of the two systems. These characteristics underscore the value of the St\"ackel transform as a classifying tool. A subject of immediate interest is to try to determine which functions $U$ give St\"ackel equivalent systems on the same manifold and also if we can generalize these results to higher dimensions. For example, recently a 3 dimensional generalization of this system was shown to be classically superintegrable for rational $k$ \cite{KKM2010JPA}. 

Finally,  it remains a further subject of research to find the closed form solutions of the second integrals of motion in the quantum case and prove that they can be chosen in a form that will admit a St\"ackel transform. This is the case for all explicitly constructed examples given in \cite{TTW2009, KKM2010quant} and in the general odd k case given in \cite{QuesneTTWodd}.

\ack{}
We thank W. Miller, F. Tremblay and A. V. Turbiner for discussions. The research of P.W. was partly supported by NSERC of Canada. 
\section*{References}
\bibliography{post2632010A}{}
\bibliographystyle{unsrt}
\end{document}